\documentclass{PoS}

\usepackage{amsmath}
\usepackage{amsfonts}
\usepackage{amssymb}
\usepackage{graphicx}

\usepackage{booktabs}
\usepackage{wrapfig}

\title{Challenges for Moduli Stabilization and String Cosmology near the Conifold}

\ShortTitle{Challenges for Moduli Stabilization and String Cosmology near the Conifold}

\author{Ralph Blumenhagen\\
        Max-Planck-Institut f\"ur Physik, M\"unchen\\
        E-mail: \email{blumenha@mpp.mpg.de}}
        
\author{\speaker{Daniela Herschmann}\\
        Max-Planck-Institut f\"ur Physik, M\"unchen\\
        E-mail: \email{herschma@mpp.mpg.de}}
        
\author{\speaker{Florian Wolf}\\
        Max-Planck-Institut f\"ur Physik, M\"unchen\\
        E-mail: \email{flowolf@mpp.mpg.de}}

\abstract{This brief article reviews a recently proposed scenario of moduli
stabilization constructed in the vicinity of a conifold locus in the complex structure moduli space.
We discuss typical features of moduli stabilization due to the logarithmic structure of the periods over the Calabi-Yau threefold.
Integrating out heavy moduli implies exponential terms in the
superpotential very reminiscent of non-perturbative
contributions. Ultimately, these terms can lead to an alignment of
axions potentially realizing large-field inflation.
Our goal is to investigate common issues appearing for moduli stabilization near the conifold and subsequent applications to string cosmology.
Even though this setup needs to be understood as a toy model, a closer
look at the validity of the effective theories reveals characteristic
obstacles, which are likely to occur in more serious scenarios as well.}

\FullConference{Corfu Summer Institute 2016 "School and Workshop on Elementary Particle Physics and Gravity"\\
		August 31 - September 23, 2016\\
		Corfu, Greece}

\newcommand{\eq}[1]{\begin{equation}
                     \begin{split} #1 \end{split}
                     \end{equation}}
\newcommand{\ov}{\overline}
\newcommand{\op}{\hspace{1pt}}		

\begin{document}
\section{Introduction}
\noindent
This article considers moduli stabilization at the conifold and addresses a major challenge of string phenomenology: combining moduli stabilization and large-field inflation. It turns out that their interplay leads to new limitations about consistent possibilities within string compactifications as well as limitations about effective low-energy theories derived from quantum gravity.
Let us briefly review these challenges in more detail, before we
outline the motivation and plan for this article.

Compactifying full string theory to a lower dimensional effective description, implies several massless moduli which have to be stabilized to be in agreement with experiments.
Giddings-Kachru-Polchinski (GKP) \cite{Giddings:2001yu} have proven that type IIB flux compactifications on warped Calabi-Yau threefolds correspond to solutions of the string equations of motion.
Their idea gave rise to the modern picture of a string landscape of vacua.
In GKP the axio-dilaton and complex structure moduli were fixed by
3-form fluxes, whereas the K\"ahler moduli remained massless due to a
no-scale structure. We show that, in view of having control over the
effective theory, the natural mechanism for K\"ahler moduli stabilization is the large volume scenario.

Cosmic inflation is a crucial ingredient of the standard model of
cosmology $\Lambda$CDM. Lately large-field inflation aroused a lot of
attention in the string-pheno community, as  its embedding into a UV
complete theory seems both to be  very challenging and to provide a
good question  to unravel general constraints from quantum gravity.
The supplement \emph{large-field} stands for a trans-Planckian distance in field space, which the inflaton is traversing during inflation.
Hence a slowly-rolling inflaton requires to have a relatively flat potential over a trans-Planckian scale.
It is quite appealing that string compactifications naturally provide axionic moduli as inflaton candidates, whose shift symmetry  prohibits UV corrections that would deform this potential on sub-Planckian scales.
Considering the zoo of possible potentials for axion inflation in string theory, there are basically two distinctive categories: periodic and polynomial inflaton potentials.
Polynomial potentials rely on a special mechanism known as axion monodromy \cite{Silverstein:2008sg}.
Let us already mention that, in this article we focus exclusively on
periodic inflaton potentials featuring an KNP-alignment behavior
\cite{Kim:2004rp}. 

Originally, our motivation arose purely from investigating other
distinguished points in the complex structure moduli space. To that end, we analyze the vicinity of a conifold locus and check for new characteristic features regarding moduli stabilization and large-field inflation.
Being only \emph{in the vicinity of} the conifold singularity guarantees the absence of uncontrolled warping effects, potentially invalidating the effective supergravity theory.
The special attribute of the conifold in the complex structure moduli space is the appearance of a logarithm in its periods over the holomorphic 3-form of the Calabi-Yau manifold.
As a consequence, it will turn out that the conic modulus will be most
heavy and that  after integrating it out, the logarithm induces
exponential terms in the superpotential, that are reminiscent of
instanton corrections.

With the  more detailed analysis being  performed in \cite{Blumenhagen:2016bfp}, 
in this proceeding article we intend to change the interpretation of
the results more towards  potential challenges of moduli stabilization
at the conifold.
In fact, the absence of warping effects and/or  a completely
controllable mass hierarchy will require strict constraints on  moduli
stabilization. These will also show up for the concrete model of  aligned inflation, afterwards.
Note that recently models of large-field inflation in string theory
have become   under pressure and skeptical examination.
The reason is that, although several nice mechanisms giving rise to
trans-Planckian field excursions have been invented, typically an
embedding into a fully-fledged setup including a proper treatment of
moduli stabilization fails due to control issues.
However, the the manifestation of such issues can be  of different nature, which makes it intricate to derive a general statement.
Nevertheless, underlying principles of quantum gravity have been proposed, such as the \emph{Weak Gravity Conjecture} (WGC) \cite{ArkaniHamed:2006dz,Rudelius:2014wla}, elucidating the global obstacles of large-field inflation.
Lately, it was also pointed out that further trouble might occur due to mass scales becoming exponentially light for moduli moving trans-Planckian distances. This was termed (extended) \emph{Swampland Conjecture} \cite{Ooguri:2006in,Baume:2016psm,Klaewer:2016kiy,Valenzuela:2016yny,Blumenhagen:2017cxt}, but shall not be part of the discussion here.

The outline for this article is therefore as follows:
we start in section 2 with introducing the theoretical concepts of flux compactifications of type IIB string theory on Calabi-Yau threefolds and derive the periods close to the conifold locus.
The typical challenges near the conifold are divided in two sections: Section 3 deduces a constraint from warping effects and self-consistency of stabilizing the heavy conic complex structure modulus.
Section 4 analyzes the validity of applying this conifold model to aligned inflation.
We conclude with a brief summary of results in section 5.

\section{Type IIB Orientifolds and Periods for the Conifold}
\noindent
The model constructed in this article is based on the mirror manifold of the Calabi-Yau threefold $\mathbb{P}_{11226}[12]$. Before we move on to its periods near the conifold, let us introduce general concepts of orientifold compactifications and the resulting moduli fields.

We start with compactifying type IIB string theory on orientifolds of Calabi-Yau threefolds $\mathcal M$, that are equipped with a holomorphic 3-form $\Omega_3$.
The orientifold projection 
 $\Omega_{\rm P} (-1)^{F_{\rm L}} \sigma$ contains
a holomorphic involution $\sigma:{\cal M}\to {\cal M}$, the world-sheet parity operator $\Omega_{\rm P}$ as well as the left-moving fermion
number $F_{\rm L}$. We choose the involution  $\sigma$ to act on the 
K\"ahler form $J$ and the holomorphic $(3,0)$-form $\Omega_3$ of the Calabi-Yau threefold $\mathcal M$ as
\eq{
  \label{op_01}
        \sigma^*: J\to J\,,\hspace{50pt} \sigma^*:\Omega_3\to-\Omega_3\,.
}
The fixed loci of this involution correspond to O$7$- and O$3$-planes, which 
in general require the presence of D$7$- and D$3$-branes  to satisfy the tadpole cancellation
conditions.
The holomorphic involution $\sigma$ of the orientifold projection splits the cohomology into even and odd parts
\eq{
 H^{p,q}(\mathcal M) = H^{p,q}_+(\mathcal M) \oplus H^{p,q}_-(\mathcal M) \,,
 \hspace{50pt}
 h^{p,q} = h^{p,q}_+ + h^{p,q}_-\,.
}
Reducing the ten-dimensional bosonic field content of type IIB string theory on the Calabi-Yau threefold $\mathcal M$ leads to numerous massless moduli in the effective four-dimensional supergravity theory \cite{Grimm:2004uq}.
The closed string moduli are summarized in table \ref{table_moduli}, where
the convention was chosen such that the imaginary parts of the moduli
correspond to axions\footnote{The full definition of the K\"ahler moduli $T_{\alpha}$ is 
given by
\eq{
\label{defkaehler}
   T_\alpha=\frac{1}{2}\op\kappa_{\alpha\beta\gamma} t^\beta t^\gamma
   +i\left(\rho_\alpha-\frac{1}{2}\op\kappa_{\alpha a b} c^a b^b\right)
  -\frac{1}{4} \op e^\phi  \kappa_{\alpha a b} {G}^a (G+\ov G)^b \,,
}
where $\kappa_{\alpha\beta\gamma}$ denote the triple intersection numbers.}.
We will call the non-axionic real part of the moduli \emph{saxions}, which should not be confused with superpartners of the axions.
In the following we have redefined the axio-dilaton as $S =
s + i \, c$. Since the axionic odd moduli do not play any role in the subsequent
discussion, we choose $h^{1,1}_-=0$ and for convenience $h^{2,1}_+=0$, such that there are no additional abelian vector superfields.

\begin{table}[ht]
\centering
\renewcommand{\arraystretch}{1.3}
\begin{tabular}{@{}cl@{\hspace{1pt}}lc@{}}
  \toprule
   number & \multicolumn{2}{c}{modulus} &  name \\
  \hline
  $1$ & $S$&$=g_s^{-1}-i\op C_0$ & axio-dilaton \\
  $h^{2,1}_- (\mathcal M)$ & $U^i$&$=u^i+i\op v^i$ & complex structure\\
 $h^{1,1}_+ (\mathcal M)$ & $T_\alpha$&$=\tau_\alpha+ i \op \rho_\alpha+\ldots$ & K\"ahler\\

 $h^{1,1}_-(\mathcal M)$ & $G^a$&$=S\op b^a+i\op c^a$ & axionic odd\\
\bottomrule
     \end{tabular} 
     \caption{\small Closed string moduli in type IIB orientifold compactifications.}
      \label{table_moduli}
\end{table}

As usual, turning on non-trivial background fluxes generates a scalar
potential for moduli, such that they are stabilized at the
minimum. For an extensive review, see for instance
\cite{Grana:2005jc, Douglas:2006es,Blumenhagen:2006ci}. We will return to these fluxes in later
sections, but at first we discuss the K\"ahler potential constituting another ingredient of the $\mathcal N = 1$ scalar potential. The K\"ahler potential of the complex structure moduli space is entirely determined by the periods.

One expands the holomorphic 3-form $\Omega_3$ as follows
\eq{
  \label{exp_02}
  \Omega_3 = X^{\lambda} \alpha_{\lambda} - F_{\lambda} \op\beta^{\lambda} \,,
}
where we introduced a symplectic basis for the third cohomology of the
Calabi-Yau manifold $\mathcal M$ by
\eq{
  \{\alpha_{\lambda},\beta^{\lambda}\} \in H^3(\mathcal M) \,, \hspace{60pt}
  \lambda =0,\ldots, h^{2,1} \,.
}
The periods 
$X^{\lambda}$ and $F_{\lambda}$ of the Calabi-Yau manifold are defined via
\eq{
\label{periods_int}
                 X^{\lambda}=\int_{A^\lambda} \Omega_3\,,\qquad  
 F_{\lambda}=\int_{B_\lambda} \Omega_3\,.
}
Here  $A^\lambda,B_\lambda\in H_3({\cal M})$ denote  a basis of
Poincare dual 3-cycles.
The $X^{\lambda}$ can be considered as homogeneous coordinates
of the complex structure moduli space. Inhomogeneous coordinates
are then defined via $U^i=X^i/X^0$ with  $i = 1,\ldots,h^{2,1}$, corresponding to the moduli listed in table \ref{table_moduli}.
Since the complex structure moduli space of $\mathcal N = 1$
orientifold compactifications features  the 
powerful  structure of special (K\"ahler) geometry, the periods $F_{\lambda}$ can be expressed as derivatives 
$F_{\lambda} = \partial F/\partial X^{\lambda}$ of a prepotential $F$.

Finally, the K\"ahler potential for these chiral
superfields is given at leading order in $\alpha'$ by
\eq{ 
\label{Kpot}
	K=-\log(S+\ov S) -2\log {\cal V} -\log \Big(-i \int \Omega_3\wedge
  \ov\Omega_3\Big) \, ,
}
where ${\cal V}$ denotes the total volume of the Calabi-Yau manifold
that can be expressed in terms of the four-cycle volumes ${\rm Re}(T_\alpha)$.

The concrete form of the K\"ahler potential in terms of the moduli depends on the region of moduli space under investigation. We are interested in the vicinity of a point in the complex structure moduli space in which one of the periods vanishes.
Let us call this period $F_1 = Z$, where $Z$ denote one coordinate of the complex structure moduli space.
The point where this coordinate shrinks to zero is a singularity called \emph{conifold}.
Moreover, for a closed loop around the conifold singularity, 
the symplectic dual period undergoes a monodromy
$X^1\to X^1 + F_1$. The remaining periods should stay  finite 
at the conifold locus.
Near $F_1\sim0$, the monodromy is captured by the period over the dual cycle containing a logarithmic term
\eq{\label{hustekuchen}X^1\sim {\rm const.} + \frac{1}{2 \pi i} Z \log{Z\,.}
}
This $\log Z$-term is the origin of all new interesting applications
of the setup,  described in the following sections. It is also the
reason why we will often refer to $Z$ as the \emph{conic modulus}.
Plugging the logarithmic form of the periods into the K\"ahler metric
$G_{Z\ov Z}$, the singularity becomes visible. In fact,  the K\"ahler metric is divergent near $Z=0$ according to the relation
\eq{
G_{Z\ov Z}\sim \log{Z \ov Z} \, .
}

As an example, let us consider a geometry with two complex structure
moduli. Hence, in this article we work exclusively with the Calabi-Yau
threefold $\mathbb{P}_{11226}[12]^{(128,2)}$.
The mirror of this  manifold has two complex structure moduli, that we now call $\psi$ and $\phi$,
appearing as deformations of the hypersurface constraint
\eq{
          P=z_1^{12}+z_2^{12}+z_3^{6}+z_4^{6}+z_5^{2}-12\psi\,
           z_1\,z_2\,z_3\,z_4\,z_5-2\phi\, z^6_1\,z^6_2\,.
}
The codimension one conifold locus is at $864\,\psi^6+\phi=1$. The
periods were computed in \cite{Berglund:1993ax,Candelas:1993dm}.
Up to quadratic order, in terms of the two
special complex structure fields $Z$ and $Y$, the periods can be
expanded as
\eq{
\label{periodsExact}
       F_0&=1\,,\\
       F_1&=Z\,,\\
     F_2&=(0.46 + 0.11i) + (1.10-2.17i)Y  -0.19\,Z \\
     & -(7.34 - 14.73i)\, Y^2 +(2.71 + 1.42i)\, YZ +(0.11 - 1.69i)\,Z^2
}
and
\eq{
X^0&=(-0.04 + 0.23i)+( 1.10 +0.06i) Y + 0.17\,Z \\
&  -(7.34 + 1.83i)\,Y^2 + (0.55 + 1.42i)\,YZ + (0.11 - 0.17i)\,Z^2\,,\\
X^1&=-{1\over 2\pi i}Z\log Z + 0.18 - 0.42\,Y -  1.43 i\, Z +\ldots\,,\\[0.1cm]
X^2&=0.09 - 2.19\,Y + 14.67\,Y^2 - 2.84i\,YZ - 0.22\,Z^2\,.
} 
%
The resulting K\"ahler potential at the conifold simplifies considerably.
For the complex structure part of the K\"ahler potential at linear
order, one finds
\eq{
K_{\rm cs}&=-\log\left[- i \left( X^\lambda \ov F_\lambda - \ov X^\lambda F_\lambda \right) \right]\\[0.1cm]
&=-\log\left[{1\over 2\pi} |Z|^2 \log\left( |Z|^2\right) +A +
{\rm  Re}Y+    B \,({\rm  Re}Y)^2 +C\, |Z|^2\ldots\right]
}
with $A=0.44$ and $B=- 19.05$ and $C= -2.86$. 

The K\"ahler potential exhibits certain shift symmetries \cite{Garcia-Etxebarria:2014wla}: obviously, there is a typical axionic shift symmetry ${\rm Im} (Y) \rightarrow {\rm Im} (Y) + f_Y$ with constant $f_Y$. Furthermore, the conic modulus obeys a continuous phase symmetry $Z = e^{2 \pi i \, \theta} |Z|$ , which turns out to be broken by higher order terms to a discrete shift symmetry $\theta \rightarrow \theta + 1$.

Note that the K\"ahler potential has no linear term in the conic modulus $Z$, which would be dominating over the $\log Z$-term for small $Z$ and hence always make the discrete shift symmetry in $\theta$ apparent.
In fact, the precise form of the K\"ahler potential depends strongly
on the choice of periods which are only fixed up to a $Sp (6; \mathbb
Z)$ transformation. This setup is therefore indeed in agreement with
other approaches, e.g. \cite{Bizet:2016paj}.

\section{Constraints from Moduli Stabilization at the Conifold}
\noindent
The effective supergravity theory commonly used for stabilizing moduli, 
does not include Kaluza-Klein states and massive string excitations.
However, such modes can become light due to red-shifting of mass scales close to the conifold singularity, which would spoil the validity of the effective theory.
In this section we will see that pure warping considerations as well as the mass hierarchy after moduli stabilization imply a constraint on the moduli avoiding an improper arrangement of the scales.

\subsection{Warping}
\noindent
In order to stabilize moduli, we will turn on 3-form fluxes inducing a potential for the scalar fields. Consequently, tadpole cancellation requires local sources in the form of D3-branes. It is well known, though, that the backreaction of such a 3-form flux and of localized D3-branes on
the geometry leads to a warped Calabi-Yau metric \cite{Giddings:2001yu}
\eq{
                 ds^2=e^{2A(y)} \eta_{\mu\nu} dx^\mu dx^\nu +
                            e^{-2A(y)} \tilde g_{mn} dy^m dy^n \, ,
}
where $\tilde g_{mn}$ denotes the Ricci-flat metric on a Calabi-Yau
threefold. As proven in \cite{Candelas:1989js} the warp factor in the conifold vicinity is related to the conic modulus $Z$ via $e^{A_{\rm con}}\sim |Z|^{1\over 3}$.
However, the moduli dependence of the warp factor is a bit more involved.
Scaling the internal metric via $\tilde g\to \Omega^2 \tilde g$
describes the overall volume modulus $\mathcal V$ of the Calabi-Yau via $\Omega\sim {\cal V}^{1\over 6}$.
It was shown in \cite{Giddings:2005ff} that the string equations of motion admit
an unconstrained deformation $\Omega$ only if the warp factor
scales non-trivially
\eq{
           e^{-4A(y)}=1+ {e^{-4A_{\rm  con}}\over \Omega^4}\sim 1+
           {1\over  ({\cal V} |Z|^2)^{2\over 3}}
\,.
}
As a conclusion, warping effects can be neglected if (and only if) we satisfy the condition
\eq{
\label{ultimateCond}
                     {\cal V} |Z|^2\gg 1\,.
}
This is known as dilute flux limit meaning that the backreaction of fluxes and branes on the internal space does not significantly modify the geometry.
In this limit the intuitive picture of the geometry is clear because the physical size of the 3-cycle $A$
\eq{
             {\rm Vol}(A)={\cal V}^{1\over 2}\left|{\textstyle \int_A \Omega_3 }\right|= ({\cal
               V} |Z|^2)^{1\over 2}
}
remains large even when one approaches the conifold singularity by decreasing $|Z|$. Hence, the constraint ${\cal V} |Z|^2\gg 1$ implies to be not arbitrarily close to the singularity, but to keep a distance where warping effects can be safely neglected.
As already mentioned, for strong warping, Kaluza-Klein modes localized in the throat 
are  red-shifted  so that their masses might become smaller than some moduli masses.
Therefore, only in this limit  one may fully trust the usual effective
low-energy supergravity theory for the moduli of the
Calabi-Yau compactification. 

Let us now investigate moduli stabilization near the conifold, i.e. for small complex structure $Z$, dedicating special emphasis on the new features that are absent in the large complex structure regime.

\subsection{Moduli Stabilization}
\noindent
The peculiarity of moduli stabilization at the conifold in contrast to other points in moduli space, becomes apparent when turning on the $\log Z$-period in eq. \eqref{periodsExact}.
Therefore, we restrict our analysis for now to the stabilization of axio-dilaton and complex structure moduli and comment on the K\"ahler moduli later. 

The R-R and NS-NS 3-form fluxes of type IIB string theory denoted by $F = d C_2$ and $H = d B_2$, respectively, give rise to the well-known Gukov-Vafa-Witten superpotential \cite{Gukov:1999ya,Taylor:1999ii}
\eq{
  \label{s_pot_02}
  W\, = \, \int_{\mathcal M} \bigl( F +i S\, H \bigr) \wedge \Omega_3  \, 
  = \, - \left( \mathfrak{f}_\lambda X^\lambda - \tilde{\mathfrak{f}}^\lambda F_\lambda \right) 
  + i S \left(  \mathfrak{h}_\lambda X^\lambda - \tilde{\mathfrak{h}}^\lambda F_\lambda \right) \, ,
}
where we used the periods \eqref{periods_int} and the quantization of the 3-form fluxes (with 3-cycles $A^\lambda,B_\lambda\in H_3({\cal M})$)
\eq{
\label{3flux}
	\frac{1}{2 \pi}\, \int_{A^\lambda} F &= \mathfrak{\tilde f} \in \mathbb Z\, ,
	\qquad \frac{1}{2 \pi}\, \int_{B_\lambda} F = \mathfrak{f} \in \mathbb Z  \\
	\frac{1}{2 \pi}\, \int_{A^\lambda} H &= \tilde h \in \mathbb Z\, ,
	\qquad \frac{1}{2 \pi}\, \int_{B_\lambda} H = h \in \mathbb Z \, .
}
Since the superpotential $W$ does so far not depend on the K\"ahler moduli, the induced scalar potential of ${\cal N} =1$ supergravity is of no-scale type\footnote{For an overall K\"ahler modulus ${\cal V} = (T + \ov T)^{3/2}$ the no-scale relation is $G^{T \ov T} D_T W \, D_{\ov T}\ov W = 3 |W|^2$ and leads immediately to the scalar potential \eqref{Vpot}.} 
\eq{
\label{Vpot}
     V=e^K \, G^{M\ov N} \, D_M W \, D_{\ov N}\ov W \, ,
}
where the indices run over the axio-dilaton as well as the complex structure moduli. For our example $\mathbb{P}_{11226}[12]^{(128,2)}$ the moduli are $M, N \in \{ S,Z,Y \}$.
The moduli are stabilized at
Minkowski minima of the scalar potential, which are determined by the F-term conditions $F_M=D_M W = \partial_M W + (\partial_M K) W=0$.

Inspired by GKP \cite{Giddings:2001yu}, let us turn on 3-form fluxes to obtain the following superpotential
\eq{
\label{superpotchoice}
	W \, =
	\, &f \left( -{1\over 2\pi i}\, Z\, \log Z + 0.18 - 0.42\,Y -  1.43 i\, Z +\ldots\, \right)\\
	&+\,  i \left( h\, S + \hat f \, Y \right) \, Z
	-\, i h' \, S -\, i \hat f' \, Y +\, \ldots	 \, ,
	}
where $f= - \mathfrak f_1$ and $h$, $\hat f$, $h'$, $\hat f'$ are constituted by 3-form fluxes summarized in \eqref{3flux} as well as numerical prefactors in the periods \eqref{periodsExact}.
Here, we allow us to be more flexible with the values of the prefactors in the superpotential than the actual Calabi-Yau threefold $\mathbb{P}_{11226}[12]$ may permit.

Up to terms vanishing in the limit $Z \rightarrow 0$, the F-term condition $F_Z = 0$ is satisfied for
\eq{
\label{Zmin}
	Z \, \sim \, \hat C \, \exp \left[ - \frac{2 \pi}{f} \, \left( h\, S + \hat f \, Y \right)  \right] \, , 
	\qquad {\rm with} \qquad
	\hat C \sim e^{7.98} \, .
	}
Thus, the conic modulus $Z$ is indeed fixed near the conifold singularity for a sufficiently large exponent, that is, $Z$ is exponentially small at the minimum.

Let us in addition compute the mass of the conic modulus assuming that it is exponentially larger than the other moduli.
At leading order the masses of the two real scalars in $Z$ are degenerate and we find
\eq{
	 M^2_Z \, = \, \frac{1}{2} G^{Z \ov Z} V_{Z \ov Z} \, \sim \,
	 \frac{M^2_{\rm Pl}}{{\rm Re}(S) {\cal V}^2 |Z|^2} \frac{f^2}{\log^2 (|Z|^2)}
	 \, \sim \, {M_{\rm s}^2\over {\cal V} |Z|^2 } \frac{f^4 g_s^{\frac{5}{2}}}{h^2} \, ,
	 }
where we used the string scale $M_s^2 = \frac{M^2_{\rm Pl} \, g_s^{1/2}}{\cal V}$.
For our effective theory to be valid, the conic mass has to be smaller than the string scale. This leads again to the condition ${\cal V} |Z|^2\gg 1$ (assuming the flux $h$ to be of order ${\cal O} (1)$).
Hence, self-consistency of our effective supergravity theory forces us already to stay in the regime of negligible warping.
     
\subsection{K\"ahler Moduli Stabilization - The Conic LVS}
\noindent
In this section,  we briefly discuss K\"ahler moduli stabilization in the vicinity of the conifold.
The condition ${\cal V} |Z|^2\gg 1$ together with the fact that the conic modulus is stabilized at exponentially small values imply a natural candidate: the large volume scenario (LVS) \cite{Balasubramanian:2005zx}.
Let us take a look if this moduli stabilization procedure is indeed fulfilling the constraint.
We consider a swiss-cheese 
Calabi-Yau  threefold with the $\alpha'$-corrected K\"ahler potential
\eq{
                K=-2\log \left( \tau_b^{3\over 2}-\tau_s^{3\over 2}
                  +{\xi\over 2}  {\rm Re}(S)^{3\over 2} \right)\,.
}
The conic large volume scenario  is then based on the superpotential
\eq{
          W_{\rm inst}(T_s)=W_0 +A_s\, Z^N\, e^{- a_s T_s}\,.
}
The Pfaffian $A_s$ is in general an unknown function of complex structure moduli.
We are interested in a large volume scenario close to the conifold point and therefore parametrize the unknown Pfaffian $A_s(Z,U)$ as $A_s(U) \times Z^N$ to extract the dependence on the conic modulus.
We do this to see under which conditions the relation ${\cal V} |Z| \gg 1$ can be satisfied.\\
The only difference to the standard LVS is the general unknown Pfaffian $A_s \rightarrow A_s\,Z^N$.
Up to Calabi-Yau geometry dependent coefficients of order one, 
after freezing the axion
$\rho_s$, the dominant terms in the scalar potential read 
\eq{
\label{ScalarPotLVS}
    V_{\rm LVS}(T)=e^{K_{cs}} {g_s\over 2}\Bigg({|a_s A_s Z^N|^2 \sqrt{\tau_s}\, e^{-2a_s \tau_s}\over {\cal V}} -
    {W_0\, |a_s A_s Z^N| \,\tau_s\, e^{-a_s \tau_s}\over {\cal V}^2} +{\xi\,
      W_0^2\over g_s^{3\over 2} \,{\cal V}^3}\bigg)\,.
}
Recall that the masses of the fields are
\eq{
\label{lvs_mass}
                M^2_{\tau_b}&\sim O(1)  {W_0^2\, \xi\over g_s^{1\over
                    2}\, {\cal V}^3} M_{\rm pl}^2\,, \qquad M_{\rho_b}^2 \sim 0\,,\\
               M^2_{\tau_s}&\sim  M^2_{\rho_s}\sim O(1) {a_s^2\, W_0^2\,  \xi^{4\over
                   3}\over g_s\, {\cal V}^2} M_{\rm pl}^2 \, ,
}
and do not depend on the parameter $A_s$. Hence, they do not depend on the conic modulus $Z$.
The small K\"ahler modulus and the volume at the minimum are stabilized at
\eq{
      \tau_s={(4\xi)^{2\over 3}\over g_s} \,,\qquad
      {\cal V}={W_0\, \xi^{1\over 3} \over 2^{1\over 3}\, g_s^{1\over
          2} |a_s A_s Z^N|}
        e^{a_s \tau_s}\,.
}
Now, the constraint ${\cal V} |Z|^2 \gg 1$ can be explicitly calculated and in principle easily be satisfied by having large $N$ or an appropriate flux choice.
We conclude that the large volume scenario is indeed the natural
procedure for K\"ahler moduli stabilization near the conifold. Any
other scheme\footnote{Trying to stabilize the K\"ahler moduli in the conifold regime with non-geometric fluxes \cite{Blumenhagen:2015kja} has usually turned out to be in serious conflicts with the mass hierarchy.} must also dynamically freeze the volume at exponentially large values.

\section{Constraints from Aligned Inflation at the Conifold}
\noindent
It was shown \cite{Blumenhagen:2016bfp} how the conifold setup
introduced above can in principle give rise to aligned inflation realizing a trans-Planckian axion field range.
In this article we will critically re-evaluate potential issues of this model with respect to three aspects:
validity of the effective theory by checking the mass hierarchies, formal constraints of string theory by taking the full periods into account and the Weak Gravity Conjecture.
Our computations are of course model dependent, but nevertheless we intend to stress that a semi-complete scenario for large-field inflation might be far from well-controlled.
The alignment model at the conifold demonstrates the increased
difficulties at one side of the hierarchy if improving the other side,
thus a typical lose-lose situation.

\subsection{Aligned Inflation}
\noindent
Let us briefly review the appearance of the alignment mechanism in the model at hand.
For that purpose recall the superpotential in eq. \eqref{superpotchoice}.
To stabilize the axio-dilaton and the second complex structure modulus $Y = y + i \xi$, we will proceed in two steps: First, we freeze all real moduli except one axionic combination.
The second step is to generate a cosine potential for the unfixed
axion giving rise to inflation. Since the inflaton is a linear
combination  of two axionic moduli entering the superpotential with different fluxes, we will encounter a novel alignment mechanism.
Of course, to justify this procedure, one has to check the hierarchy
of  the mass scales, a posteriori. Only if we manage to establish a proper mass hierarchy, heavier moduli may securely be integrated out and our effective two step approach is valid.

A large mass hierarchy allows us to 
integrate out $Z$, such that the superpotential \eqref{superpotchoice} simplifies to
\eq{
\label{superpot0}
 	W_{\rm eff} = i \, \left( f \, + \,  h\, S \, +\, \hat f \, Y \right) \, + \, \ldots \, ,
 	}
where the dots denote higher order polynomial terms  like $Z^n$ or $Y^n$ which are all subleading in the vicinity of the conifold $Z,Y \rightarrow 0$. We will comment on these terms again in a subsequent section.
Note that this superpotential is entirely tree-level.

Integrating out $Z$ reduces also the K\"ahler potential \eqref{Kpot} to a simple effective expression.
One can show that the necessity of staying inside the physical domain,
i.e. having $g_s^{-1} = {\rm Re} (S) > 1$, demands to include also
a term of quadratic order in $Y$ in the K\"ahler potential. For simplicity we introduce the parameter $\kappa \in \mathbb R$ and use the K\"ahler potential
\eq{
	K_{\rm eff} \, =\, -\log(S+\ov S) -2\log {\cal V} -\log \Big( A + \kappa \, {\rm Re} (Y) + {\rm Re} (Y)^2 \Big) \, .
	}
Solving the F-term conditions $D_S W = D_Y W= 0$ yields a Minkowski minimum and for $ f/h'\gg 1$, $\kappa<1$ the values of
the moduli lie in the perturbative regime of  small string coupling as well as
small complex structure modulus $Y$.
The orthogonal combination to $(\hat f' \, \zeta + h'\, c)$ is axionic
and still massless at this step. 

Let us define the axionic combination $\Theta = (h c + \hat f \xi)$ as
the inflaton field, which originates from $\Theta = {\rm arg} (Z)$ according to \eqref{Zmin}. This choice does not correspond to the axion, orthogonal to the state $(\hat f' \, \zeta + h'\, c)$ as pointed out above, but was chosen for convenience since the inflaton will be stabilized by a term proportional to $Z$.

Integrating out the heavy conic modulus $Z$ reduces the superpotential \eqref{superpotchoice} not only to the tree-level terms in \eqref{superpot0}. Instead the full expression reads
\eq{
\label{superpot1}
 	W_{\rm eff} &= i \, \left( f \, + \,  h\, S \, +\, \hat f \, Y \right) \, + 
 	\frac{f \hat C}{2 \pi i}  \exp \left[ - \frac{2 \pi}{f} \, \left( h\, S + \hat f \, Y \right)  \right] \\
 	&= i \, \left( f \, + \,  h\, S \, +\, \hat f \, Y \right) \, + 
 	\frac{f \hat C}{2 \pi i} \, |Z| \,   \exp \left[ - \frac{2 \pi}{f} \, \Theta  \right]\, .
 	}
The additional term in the superpotential will freeze the inflaton $\Theta$ and generate its potential. Remarkably, the exponential term has the same form as coming from an instanton, although it was induced by fluxes. This represents a main feature of this setup and relies on integrating out the conic modulus. It remains to be seen whether this contribution is related to true instantons on the mirror dual type IIA side.

In order to stabilize the inflaton $\Theta$, one needs to find a new minimum of the scalar potential determined by the entire superpotential \eqref{superpot1}.
Taking the backreaction of the exponential term on the stabilization of the saxions into account,
the potential for the canonically normalized axion $\tilde\Theta={h'\over \sqrt{A} \hat f'} \Theta$ reads
\eq{
\label{pot_inflaton}
	V_{\rm eff} \, = \, {4 |Z|^2\,\over
      A {\cal V}^2}\ {f h^2 \over h'}\,
    \Big(1 - \cos\big({\textstyle {2\pi \sqrt{A} (h \hat f' - h' \hat f) \over f h'}}\tilde \Theta\big)\Big) \,
     \equiv \, V_0 \Big( 1 - \cos \big( {\textstyle \frac{\tilde \Theta}{f_{\tilde \Theta}}} \big) \Big) \, .
	}
%
%
%
%
%
%
%
As expected, the potential exhibits now a Minkowski minimum at $\Theta=0$.
The axion decay constant $f_{\tilde \Theta}$ reads
\eq{
	f_{\tilde \Theta} = \frac{f}{2 \pi \sqrt{A}} \, \frac{h'}{h \hat f' - h' \hat f} \, .
	}
Obviously, one is able to achieve a trans-Planckian axion decay constant, simply by alignment the fluxes as $(h \hat f' - h' \hat f)<h'$.
This is very reminiscent of the KNP-axion alignment mechanism \cite{Kim:2004rp}, with the main difference
that one linear combination of axions is fixed by fluxes at linear
order in the fields and only the second combination by instanton-like
terms.

\subsection*{Corrections to the superpotential}
\noindent
Great care has to be taken of higher order polynomial terms in the periods as they contribute to the superpotential \eqref{superpot1} and might be dominating over the exponential term.
Of course, terms like $Z^n$ with $n \ge 2$ will be subleading since they are exponentially suppressed, but this is not necessarily true for terms like $Y^n$. It turns out that numerical prefactors governed by the underlying geometry, decide whether large-field inflation can occur.

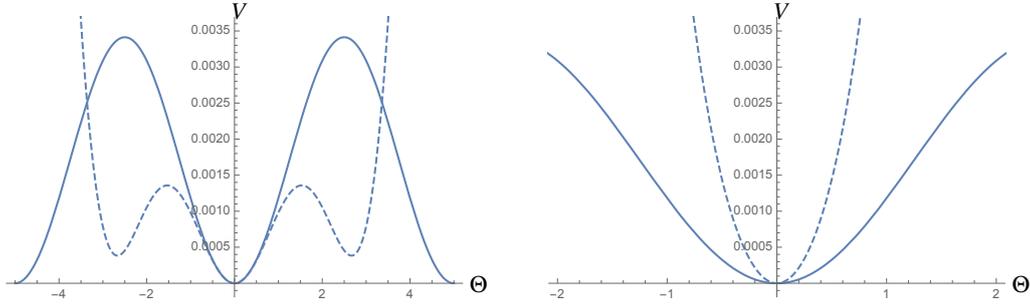
\begin{figure}[!ht]
 \centering
\includegraphics[width=0.4\textwidth]{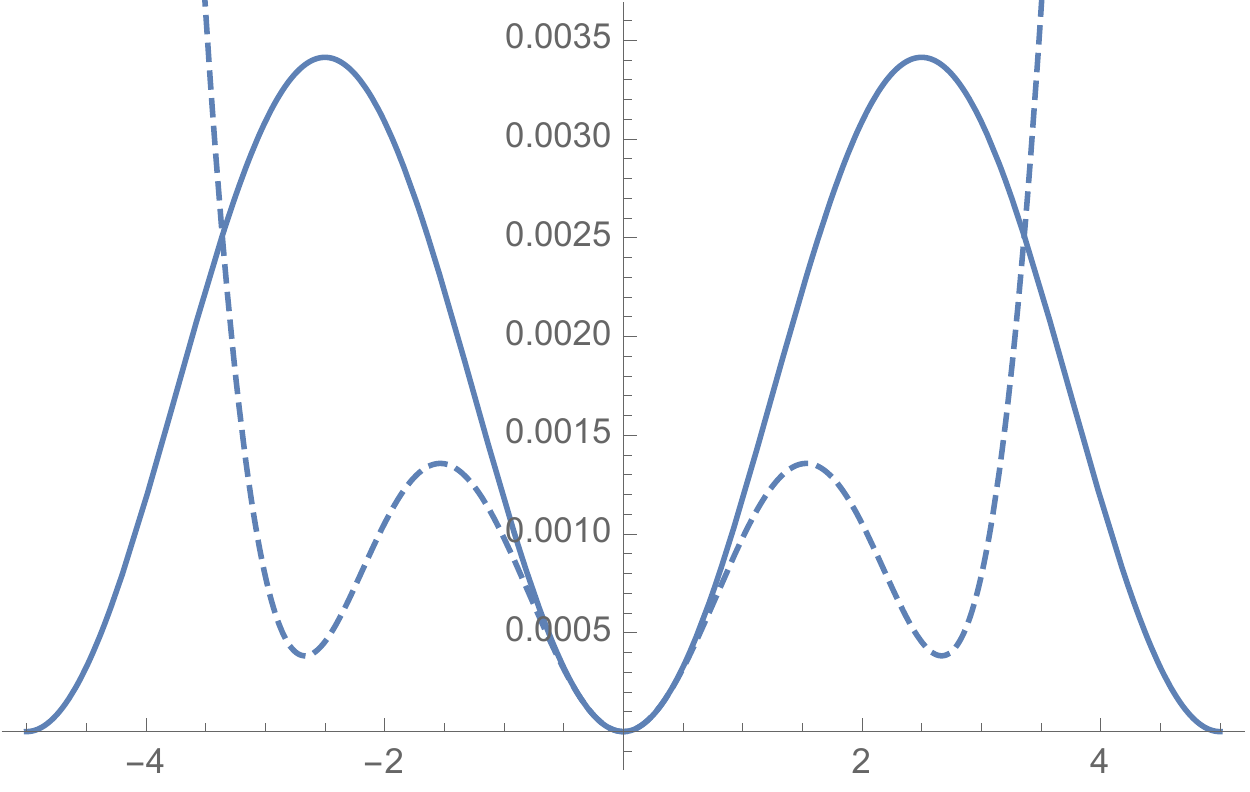}
 \begin{picture}(0,0)
      \put(-90,108){\footnotesize $V$}
    \put(-,5){\footnotesize $\Theta$}
  \end{picture}
\hspace{0.8cm}
\includegraphics[width=0.4\textwidth]{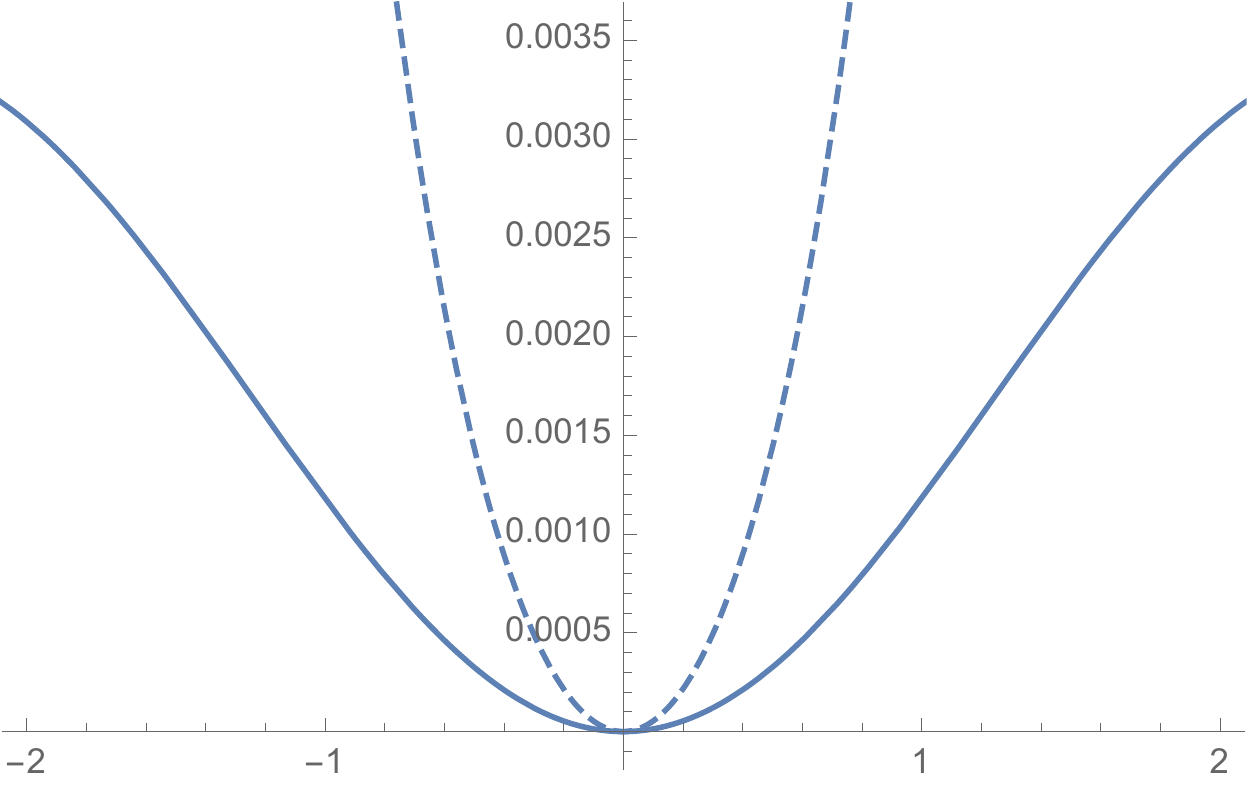} 
\begin{picture}(0,0)
   \put(-90,108){\footnotesize $V$}
   \put(-1,5){\footnotesize $\Theta$}
  \end{picture}
\caption{\label{fig_D} Scalar potential (dashed lines) for the  axion 
  $\Theta$ for $f=10$, $h'=\hat f'=1$, $h=-\hat f=1$, $\hat C=1$,
  $A=0.1$, $B=0.01$ in the left-handed  plot and $B=0.1$ in the
   right-handed  plot. For comparison, the solid lines show the potential for $B=0$.
   Only for $B=0.01$ the new potential agrees with the old one for some field distance and large-field inflation is possible.
}
\end{figure}

Let us be more precise and include the term $i B Y^2$ in the superpotential \eqref{superpot1}
\eq{
\label{finalchallengeb}
         W_{\rm eff}=i\alpha\big( f+  h'\,    S + \hat f'\,  Y\big) +iB Y^2
         +{f \hat C\over 2\pi i} \, \exp\!\Big(\! -{\textstyle {2\pi \over f}} \big(h S
              + \hat f Y\big)\Big)\, .
}
In fact, our effective theory of aligned inflation is only true if the prefactor $B$ is small enough as demonstrated in figure \ref{fig_D}.

\subsection{Mass Hierarchy}
\noindent
\begin{wrapfigure}{R}{0.5\textwidth}
\centering
\renewcommand{\arraystretch}{2.0}
\begin{tabular}{cc}
  \toprule
   Scale & $(\text{Mass})^2$ in $M^2_{\rm Pl}$\\
 \hline
  string scale $M^2_{\rm s}$ & $\displaystyle
  \frac{1}{f^{1/2}\, {\cal V}}$ \\
  Kaluza-Klein scale $M^2_{\rm KK}$ & $\displaystyle \frac{1}{{\cal V}^{4/3}}$ \\
  conic c.s. modulus $M^2_Z$ &
  $\displaystyle 
     {f \over {\cal V}^2 |Z|^2 }$ \\
  inflationary mass scale  $M^{2}_{\rm inf}$ & $\displaystyle 
   \frac{f^{1/2} |Z|}{\mathcal V}$\\
    other moduli
  $M^2_{\rm mod}$ & $\displaystyle  \frac{f}{\mathcal V^2}$ \\
    large K\"ahler modulus $M^2_{\tau_b}$ &
  $\displaystyle \frac{ f^{5/2}}{ \mathcal{V}^3 }$  \\
  inflaton $M^2_{\tilde \Theta}$ & $\displaystyle
   \frac{|Z|^2}{f \, \mathcal V^2}$\\[4mm]
\bottomrule
     \end{tabular} 
    \caption{\label{table_B}  Moduli masses and scales with $g_s\sim 1/f$.}
\end{wrapfigure}

To briefly recapitulate our approach, we had started with an effective supergravity ansatz and then integrated out successively heavy moduli. This procedure was a priori not justified and is only applicable for a proper mass hierarchy.
We will analyze the mass scales now explicitly, but drop numerical prefactors and fluxes of order $\mathcal O (1)$ (not the flux $f$) since we are only interested in parametrical control.

The masses and scales as listed in table \ref{table_B}, depend mainly on the large overall volume ${\cal V}$, the exponentially small conic modulus $Z$ and a global flux $f$.
It is crucial to consider also the string scale $M^2_{\rm s} \sim g_s^{1/2}/{\cal V}$, Kaluza-Klein scale $M^2_{\rm KK}$ and the inflationary energy scale $M^2_{\rm  inf} \sim V_0^{1/2}$.
Note that we abbreviated the masses of the other moduli given by $(M^2_{\rm mod})^i_{\ j} = \frac{1}{2} G^{ik} \partial_k \partial_j V $.

Let us be more specific about the hierarchy displayed in table \ref{table_B}. For that purpose, recall the important constraint ${\cal V} |Z|^2\gg 1$ derived in \eqref{ultimateCond}.
In fact, at a first glance the hierarchy looks quite promising: on the one hand single-field inflation is easy to assure because the inflaton is exponentially suppressed relative to the other moduli stabilized at tree-level. On the other hand, the usage of all the effective theories can be justified since the conic modulus $Z$ is heavier than the other moduli, but lighter than string and Kaluza-Klein scales as necessary. In addition, the inflationary mass scale $M_{\rm inf}$ is not automatically too high, which is a commonly faced issue for models of large-field inflation.

The drawback comes from the K\"ahler moduli. Since they get stabilized via small non-perturbative effects, it is much more subtle to integrate them out and achieve an effective theory solely for the inflaton.
Indeed, this problem is reflected by the mass hierarchy if we naively assume the K\"ahler moduli to be fixed via the large volume scenario \cite{Balasubramanian:2005zx}. 
The large 4-cycle modulus is lighter than the inflaton,
i.e. $M_{\tau_b} < M_{\tilde \Theta}$, and therefore violating
single-field axion inflation. This control issue is not just a 
short-coming of this specific model, but is consistent with the
appearance of serious control issues once one tries to embed promising
looking schemes of realizing trans-Planckian field excursions into
fully fledged string compactifications.

\subsection{Weak Gravity Conjecture}
\noindent
Let us briefly comment on applying the WGC to our conifold setup.
In its original form the WGC states that gravity is the weakest force, which can be transferred to an instanton with action $S_{\rm inst}$ coupling to an axion with decay constant $f_{\rm inst}$ as follows
\eq{
	 S_{\rm inst} \, f_{\rm inst} \le 1 \, .
	 }
Although the exponential term in the superpotential \eqref{superpot1} does only appear after integrating out other moduli, it is still instanton-like and one can check if it satisfies the WGC.

It turns out that we indeed fulfil the \emph{mild} form of the WGC, meaning that there exists \emph{some} instanton in agreement with the conjecture above.
To prove this, let us consider for instance a $D(-1)$-instanton with action and decay constant given by
\eq{
             S_{D(-1)}={2\pi s_0}={2\pi f \over h'}\,,\qquad
          f_{D(-1)}={h'\over 2\pi \sqrt{A} \hat f'}={f_{\tilde\Theta}\over k} \, ,
}
with $k=(f\, \hat f')/(h \hat f' - h' \hat f)$.
Adding the $D(-1)$-instanton to our conifold setup, the inflaton potential will be modified and schematically of the form
\eq{
	V \sim  e^{-2\,S_{\rm inst}} \, \Big( 1 - \cos \big(
            {\textstyle \frac{\tilde \Theta}{f_{\tilde \Theta}}} \big) \Big) +
	e^{-2\,S_{D(-1)}} \, \Big( 1 - \cos \big(
            {\textstyle \frac{k \, \tilde \Theta}{f_{\tilde \Theta}}} \big) \Big) \, .
	}
Since we have $S_{\rm
 inst}<S_{D(-1)}$ the first term is dominant and realizes
inflation.
If $k$ is sufficiently large, the axion decay constant
$f_{\tilde \Theta}/k$ can be sub-Planckian, satisfying the mild form of the WGC.  For more information about this loop-hole see \cite{Rudelius:2015xta,Brown:2015iha,Hebecker:2015rya}.

\section{Conclusions}
\noindent
This article started with discussing physical implications of moduli stabilization at the conifold. It turned out to be crucial to stay in the regime of negligible warping for consistent stabilization.
In other words, our ansatz is only valid if we work not at the singularity but in its vicinity.
There, K\"ahler moduli stabilization via the LVS satisfies the consistency constraint.
The logarithmic structure of the periods is an intrinsic property even at the vicinity of the conifold locus.
%
%
%
%
Let us also stress that, after integrating out heavy moduli, theses periods led to exponential terms in the superpotential very reminiscent to non-perturbative contributions.

This tempted us to investigate aligned inflation near the conifold.
However, taking the periods more serious implied corrections to the superpotential possibly spoiling large-field inflation.
Analyzing other potential issues, we could satisfy the Weak Gravity Conjecture, but encountered difficulties with the mass hierarchies.
So, although we started by determining a constraint to avoid uncontrolled warping effects, it remains 
delicate to construct even a toy model, where one can safely trust the validity of all effective theories.

Nevertheless, we showed that the string landscape is rich and in particular, other points in moduli space may have interesting new consequences for phenomenology such as large-field inflation. 
To build up a completely consistent scenario with alignment and/or the conifold locus is an intriguing, but challenging task left for future work.


\end{document}